%% file: charm.tex
\documentclass[12pt]{article}

\def\pbnr{}
\def\speaker{Samuel Harnew}
\def\onbehalfof{LHC$\mathrm{b}$}
\def\title{CP Violation and Mixing in Multi-body \D decays}
\def\affiliation{School of Physics and Astronomy\\
The University of Bristol, Bristol, UK}
\def\support{The workshop was supported by the University of Manchester, IPPP, STFC, and IOP}

\usepackage{upgreek}
\usepackage{xspace}
\usepackage{amsmath}
\usepackage{url}

\usepackage{placeins}
\input charmmacros.tex

\def\Journal#1#2#3#4{{#1} {\bf #2}, #3 (#4)}

\def\be{\begin{equation}}
\def\ee{\end{equation}}
\def\bea{\begin{eqnarray}}
\def\eea{\end{eqnarray}}

\def\PD      {\ensuremath{D}\xspace}                 
\def\D       {\ensuremath{\PD}\xspace}
\def\Dbar    {\kern 0.2em\overline{\kern -0.2em \PD}{}\xspace}
\def\Dz      {\ensuremath{\D^0}\xspace}
\def\Dzb     {\ensuremath{\Dbar^0}\xspace}

\newcommand{\catchyName}{complex interference parameter}

\newcommand{\cleoc}{\mbox{{C}{L}{E}{O}{-}{c}}}

\newcommand{\Dzero}{\Dz}       
\newcommand{\DzeroBar}{ \Dzb } 

\newcommand{\degrees}{\ensuremath{\mbox{}^o}}

\newcommand{\eqnref}[1]{Eq.~\ref{#1}}

\newcommand{\secref}[1]{Sec.~\ref{#1}}

\newcommand{\figref}[1]{Fig.~\ref{#1}}

\newcommand{\Imag}{\ensuremath{\mathit{Im}}}
\newcommand{\Real}{\ensuremath{\mathit{Re}}} 

\newcommand{\Z}{\ensuremath{\mathcal{Z}^f}}
\newcommand{\ImZ}{\ensuremath{\Imag \Z}}
\newcommand{\ReZ}{\ensuremath{\Real \Z}}

\newcommand{\ZKpipipi}{\ensuremath{\mathcal{Z}^{K3\pi}}}
\newcommand{\AvePDiffKpipipi}{\ensuremath{\delta_{D}^{K3\pi}}}
\newcommand{\CoFacKpipipi}{\ensuremath{R_{D}^{K3\pi}}}

\newcommand{\CoherenceFactor}{\ensuremath{R_{D}^{f}}}

\newcommand{\AveStrongPhaseDiff}{\ensuremath{\delta_{D}^{f}}}

\newcommand{\rD}{\ensuremath{r_{\mathit{Df}}}}

\newcommand{\gamt}{\ensuremath{\Gamma t}}

\usepackage{relsize}
\usepackage{units}
\usepackage{color}
\usepackage{graphicx}
\usepackage{subfigure}
\usepackage{amsmath}
\usepackage{xfrac}
\usepackage{amssymb}
\usepackage{rotating}

\numberwithin{equation}{section}

\begin{document}
\begin{titlepage}
\pubblock

\vfill
\Title{\title}
\vfill
\Author{\speaker\SupportedBy{\support}\OnBehalf{\onbehalfof}}
\Address{\affiliation}
\vfill
\begin{Abstract}

We present recent LHCb results and future prospects for CP violation and mixing measurements in multi-body charm decays. The complex amplitude structure of multi-body decays provides unique sensitivity to CP violation localised in certain phase space regions.
A model-independent search in the phase space of $D \to \pi^{+}\pi^{-}\pi^{+}\pi^{-}$ and $D \to K^{+}K^{-}\pi^{+}\pi^{-}$ decays showed no evidence for localised CP violation. If one assumes the no CP violation hypothesis, the probability of getting the observed results is $9.1\%$ and $41\%$, respectively. 

The model-independent determination of gamma from $B\to DK$ requires external input to account for the interference of $\Dzero$ and $\DzeroBar$ amplitudes to the same final state. Previously this input could only be obtained at the charm threshold, but recently it has been proposed that \D mixing can provide complimentary information. For the example of $D \to K^{+}\pi^{-}\pi^{+}\pi^{-}$ decays, it is shown that charm mixing can be used to considerably improve current constraints on the coherence factor and average strong phase difference, with existing data.


\end{Abstract}
\vfill
\begin{Presented}
\venue
\end{Presented}
\vfill
\end{titlepage}
\def\thefootnote{\fnsymbol{footnote}}
\setcounter{footnote}{0}
%

\section{Introduction}

The LHCb detector~\cite{LHCbJINST} is a single-arm forward spectrometer covering a unique pseudo-rapidity range $2 < \eta < 5$. The detector is specialised for the study of $b$ and $c$ quarks, making it ideal for measurements of CP violation (CPV) and mixing in the charm sector. Essential to the study of hadronic decay modes are two Ring Imaging Cherenkov detectors that provide particle identification, crucial for suppressing backgrounds. The tracking system provides an excellent impact parameter resolution, important for identifying heavy flavour decays at trigger level. 

Multi-body charm decays offer the opportunity to study CPV effects localised in phase space, providing sensitivity to phenomena that might get `washed out' in global decay rate asymmetries. In \secref{sec:CP} we present a search for local CPV in $D \to \pi^{+}\pi^{-}\pi^{+}\pi^{-}$ and $D \to K^{+}K^{-}\pi^{+}\pi^{-}$ decays. 

Quantum-correlated data provide important information on charm interference parameters that play a crucial role in the precision measurement of gamma from $B \to DK$ and related decays, where the details of the analysis depends on the final state of the \D~\cite{fs1, fs2,fs3,fs4,fs5,fs6}. In \secref{sec:Mixing} we discuss the possibility of constraining the $D \to K^{-}\pi^{+}\pi^{-}\pi^{+}$ coherence factor \CoFacKpipipi\ and average strong phase difference \AvePDiffKpipipi~\cite{CF} using input from \D mixing~\cite{Phen}. It is thought that such a measurement is already possible with data collected at LHCb.

\section{A search for local CP asymmetries at LHCb }

\label{sec:CP}

LHCb has performed searches for local CP asymmetries in several multi-body decay modes. Recent results on three body decays are discussed by Sam Gregson in these proceedings under the title ``Direct CP violation in the decays $D^{+} \to \phi \pi^{+}$ and $D_{s}^{+} \to K_{s}^{0}\pi^+$''. Here we present a search for local CP asymmetries in the four body $D \to \pi^{+}\pi^{-}\pi^{+}\pi^{-}$ and $D \to K^{+}K^{-}\pi^{+}\pi^{-}$ decays~\cite{cpvFourBody}. These decays are singly Cabibbo suppressed, so contain contributions from both loop and tree diagrams. The loop diagrams are particularly sensitive to new physics which may enhance CP violating effects~\cite{charmCPtheory}. 

\begin{figure}[ht]
    \hspace{0.5cm}
    \begin{minipage}[b]{0.48\linewidth}
        \centering
        \includegraphics[width=\linewidth]{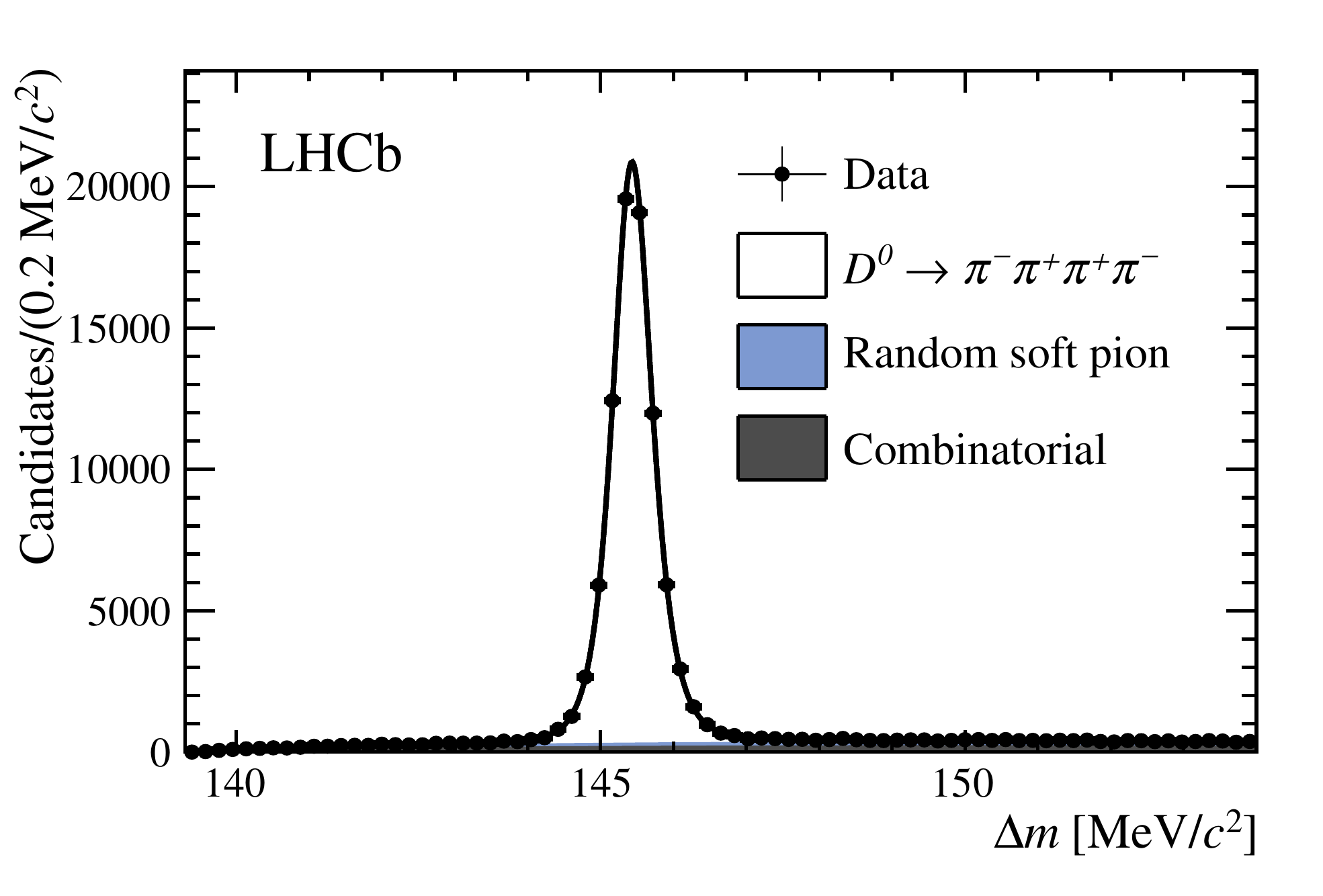}
   \end{minipage}
    \hspace{0.5cm}
    \begin{minipage}[b]{0.48\linewidth}
        \centering
        \includegraphics[width=\linewidth]{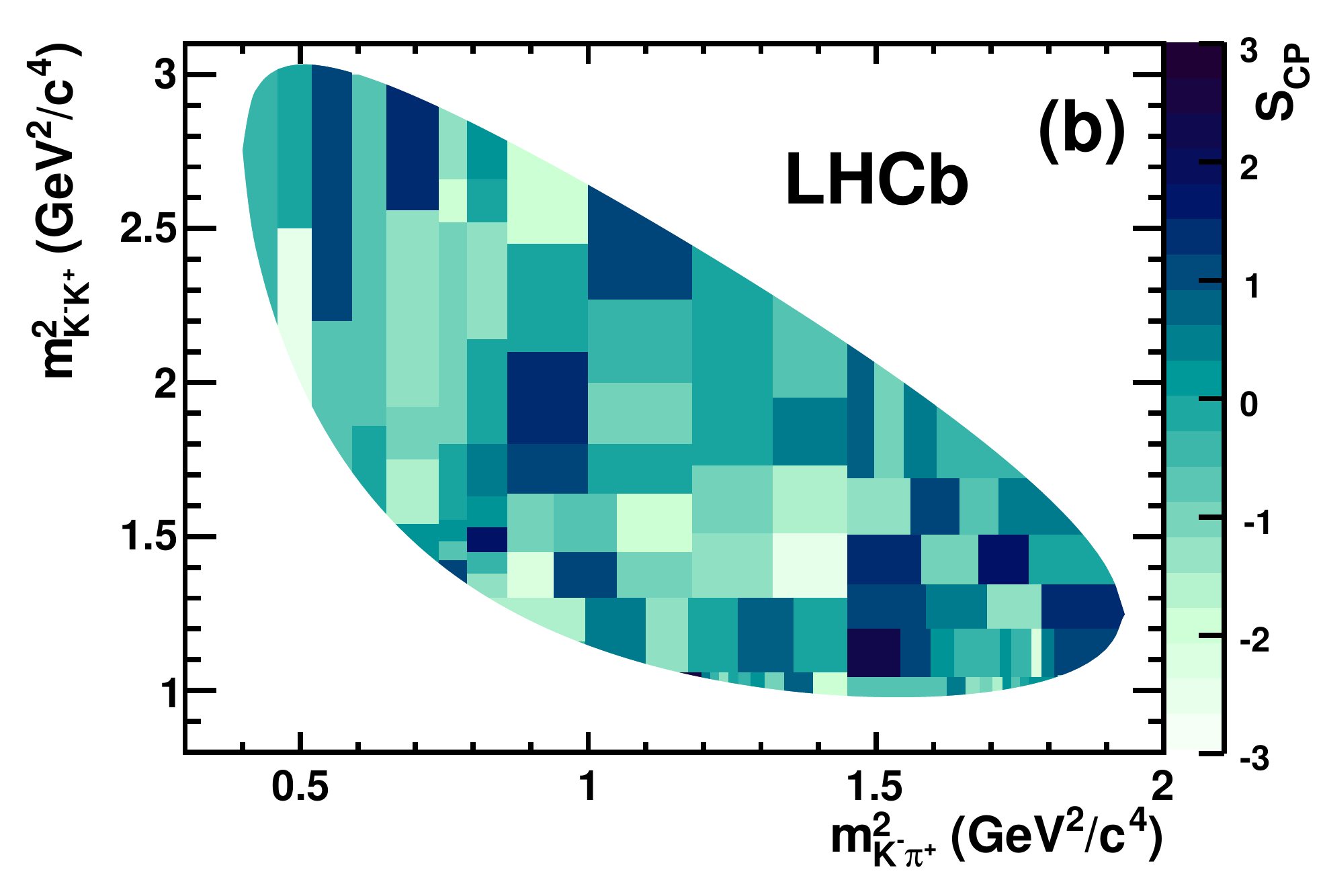}
    \end{minipage}
    \caption{ (left) $\Delta m$ projection of a 2D fit in $m(\pi^{+}\pi^{-}\pi^{+}\pi^{-})$ and $\Delta m$, superimposed with the signal candidates. Fits include various sources of backgrounds that are described in~\cite{cpvFourBody}. (right) $S_{CP}$ variation across the Dalitz plot for $\D^{+} \to K^{+}K^{-}\pi^{+}$. \label{fig:scp} }
\end{figure}

The analysis was performed using $1\mathrm{fb}^{-1}$ of data collected by LHCb during 2011.  \D mesons are reconstructed from the decay chains $D^{*+} \to \Dzero \pi_{s}^{+}$ and $D^{*-} \to \DzeroBar \pi_{s}^{-}$ where the charge of the slow pion, $\pi_{s}$, identifies the flavour of the \D meson at production. The analysis uses the `Miranda method'~\cite{miranda} which has been used in many similar searches. 
The multi-body phase space is split into $N$ independent volumes, and the variable $S^{i}_{CP}$ gives the significance of CPV in volume $i$. 
The number of $\Dzero \to f$ events in volume $i$ is given by $N_{i}(\Dzero)$ with an uncertainty $\sigma_{i}(\Dzero)$, where $f$ represents a given final state. 
The equivalent quantities for the CP conjugate process are $N_{i}(\DzeroBar)$ and $\sigma_{i}(\DzeroBar)$.
\begin{align}
 S^{i}_{CP} = \frac{ N_{i}( \Dzero ) - \alpha N_{i}( \DzeroBar ) }{\sqrt{ \alpha \left ( \sigma_{i}^{2}( \Dzero ) + \sigma_{i}^{2}( \DzeroBar ) \right ) }}  \ \ \ \ \ \  \ \ \ \ \ \ 
 \alpha = \frac{ \sum_{i} N_{i}( \Dzero ) }{ \sum_{i} N_{i}( \DzeroBar ) }
\end{align}
The quantity $\alpha$ is used to remove any global asymmetry. This includes removing sensitivity to global CPV, but also any \Dzero \DzeroBar\ production and global detection asymmetries. Describing the kinematics of four body decays requires 5 dimensions, making it difficult to visualise the variation of $S^{i}_{CP}$ across this space. \figref{fig:scp} shows the $S^{i}_{CP}$ distribution for the 3 body $\D^{+} \to K^{+}K^{-}\pi^{+}$ decay~\cite{3bodyRef}. The phase space is partitioned to give a similar number of events in each volume. A similar method is used to partition the five dimensional phase space of $D \to \pi^{+}\pi^{-}\pi^{+}\pi^{-}$ and $D \to K^{+}K^{-}\pi^{+}\pi^{-}$ decays.

\begin{figure}[ht]
    \begin{minipage}[b]{0.48\linewidth}
        \centering
        \includegraphics[width=\linewidth]{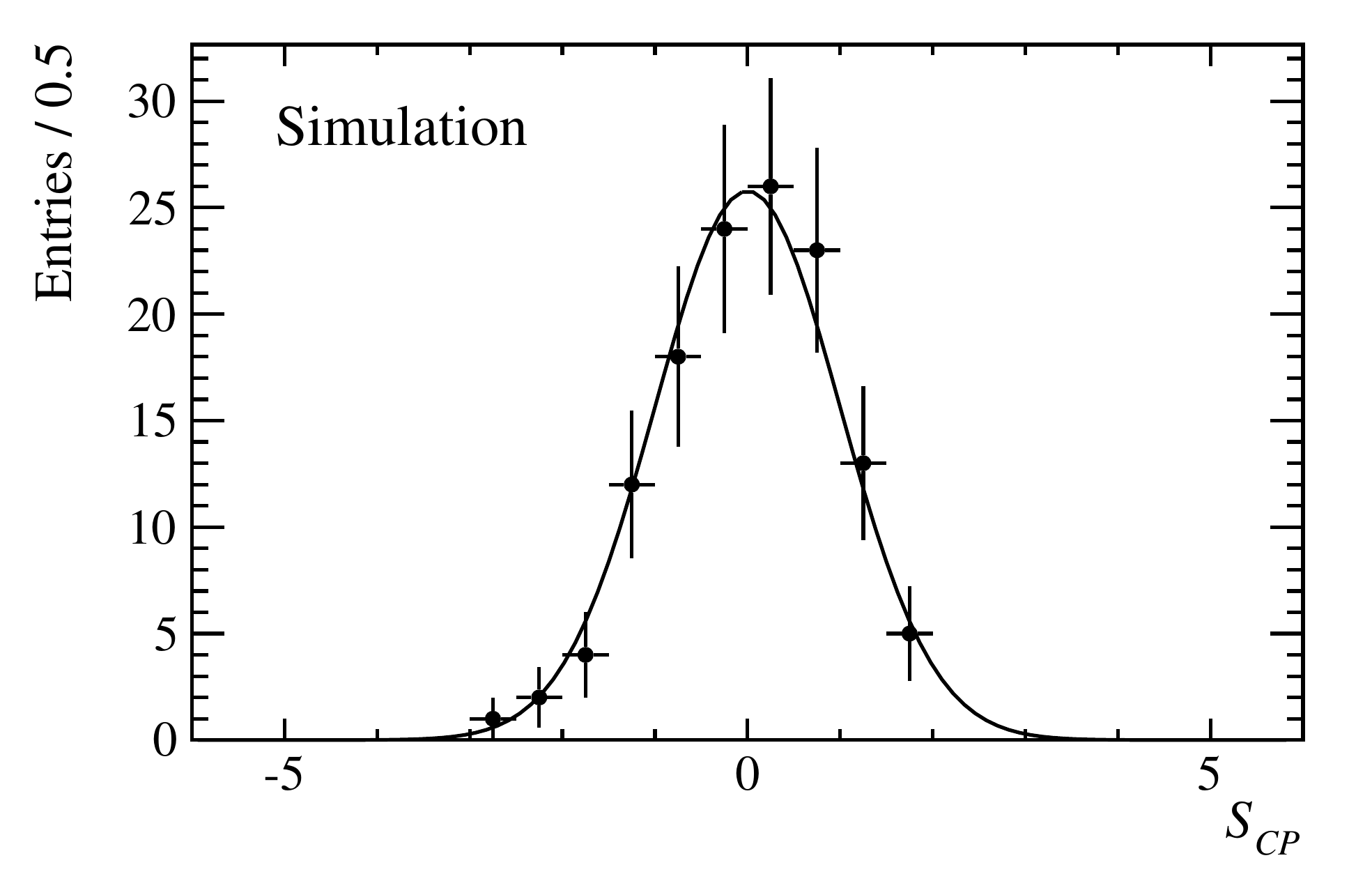}
    \end{minipage}
    \hspace{0.5cm}
    \begin{minipage}[b]{0.48\linewidth}
        \centering
        \includegraphics[width=\linewidth]{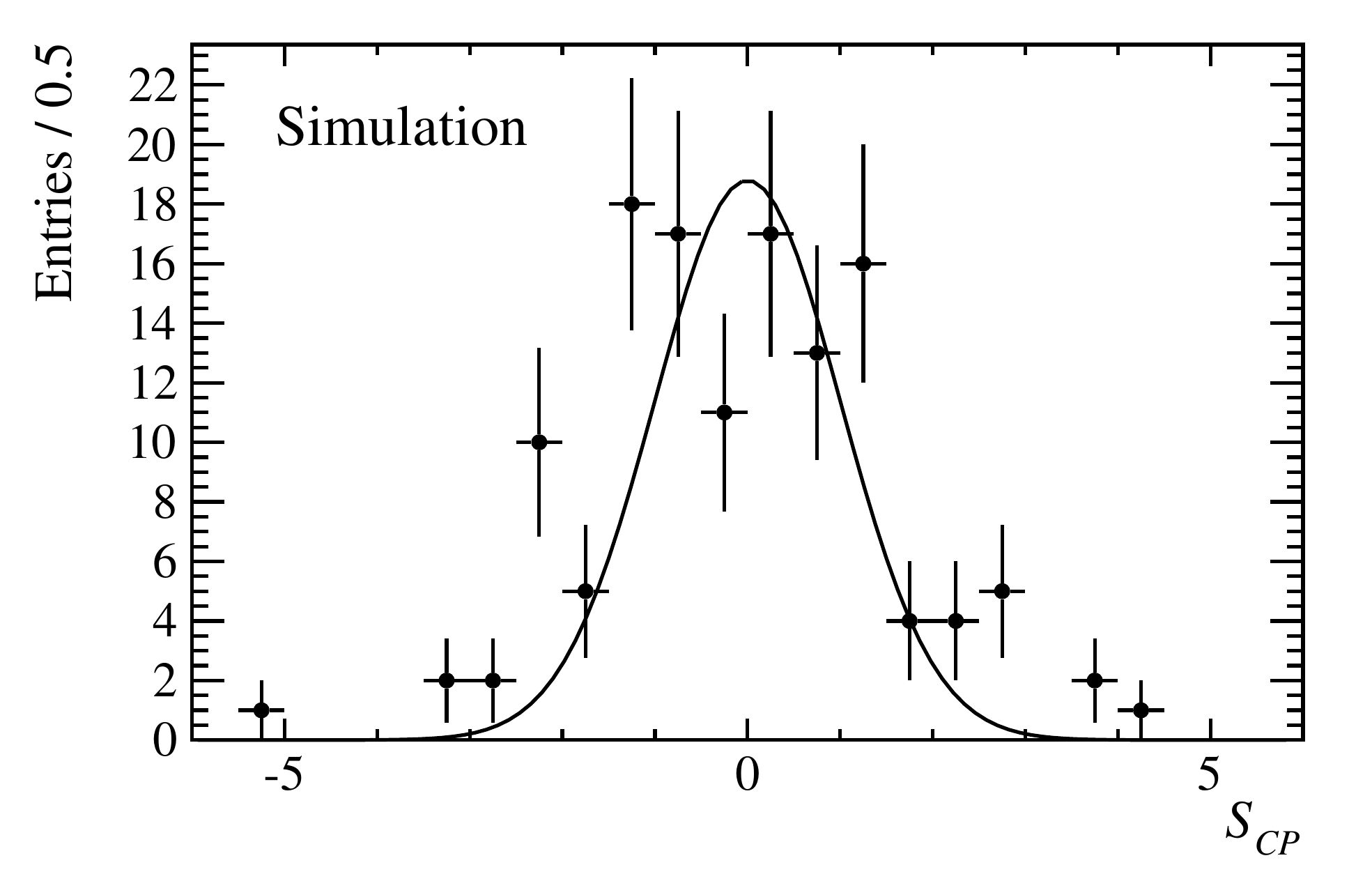}
    \end{minipage}
    \caption{The $S_{CP}$ distribution for (left) toy data generated with no CPV  (right) toy data generated with CPV. \label{toyExp}}
\end{figure}

In the case of no CPV, one would expect the $S^{i}_{CP}$ to be distributed as a gaussian, with mean 0 and unity width. To identify the presence of CPV the squared $S^{i}_{CP}$ are summed to form a $\chi^2$ statistic with $N_{bins} - 1$ degrees of freedom,
\begin{align}
\chi^{2} =  \sum_{i} \left ( S^{i}_{CP} \right )^{2},
\end{align}
from which a p-value is calculated. The p-value gives the probability of getting a larger $\chi^{2}$ than the one measured, assuming the no CPV hypothesis is true. To demonstrate the method, two sets of simulated signal events (ignoring detector resolution effects) have been generated; the first contains no local CP asymmetries, while the second contains a phase difference of $10 \degrees$ between the $\Dzero \to a_{1}(1260)^{+} \pi^{-}$ and $\DzeroBar \to a_{1}(1260)^{-} \pi^{+}$ decays. \figref{toyExp} shows the $S^{i}_{CP}$ distribution for both cases; the example with no CPV yields a p-value of $85.6\%$, whereas the example with local CPV gives $1.1 \times 10^{-16}$. 

In the analysis of LHCb data, signal yields are extracted from a 2D maximum likelihood fit in $m(hhhh)$ and $\Delta m = m(\pi_{s}hhhh) - m(hhhh)$, where $h$ represents a pion or a kaon candidate. \figref{fig:scp} shows a 1D projection of this fit for the $D \to \pi^{-}\pi^{+}\pi^{-}\pi^{+}$ channel. 
The signal yields in $D \to \pi^{-}\pi^{+}\pi^{-}\pi^{+}$ and $D \to K^{-}K^{+}\pi^{-}\pi^{+}$ decays are $330,000$ and $57,000$, respectively.  

To check for any experimental biases that could fake the presence of local CPV, the Cabibbo favoured $D \to K^{-}\pi^{+}\pi^{-}\pi^{+}$ decay is used as a control channel. In $2.9$ million signal events, no sign of any experimental bias is observed with $\chi^{2}/\mathrm{dof} = 113.4/127$ giving a p-value of 80.0\%. 

\begin{figure}[ht]
    \begin{minipage}[b]{0.48\linewidth}
        \centering
        \includegraphics[width=\linewidth]{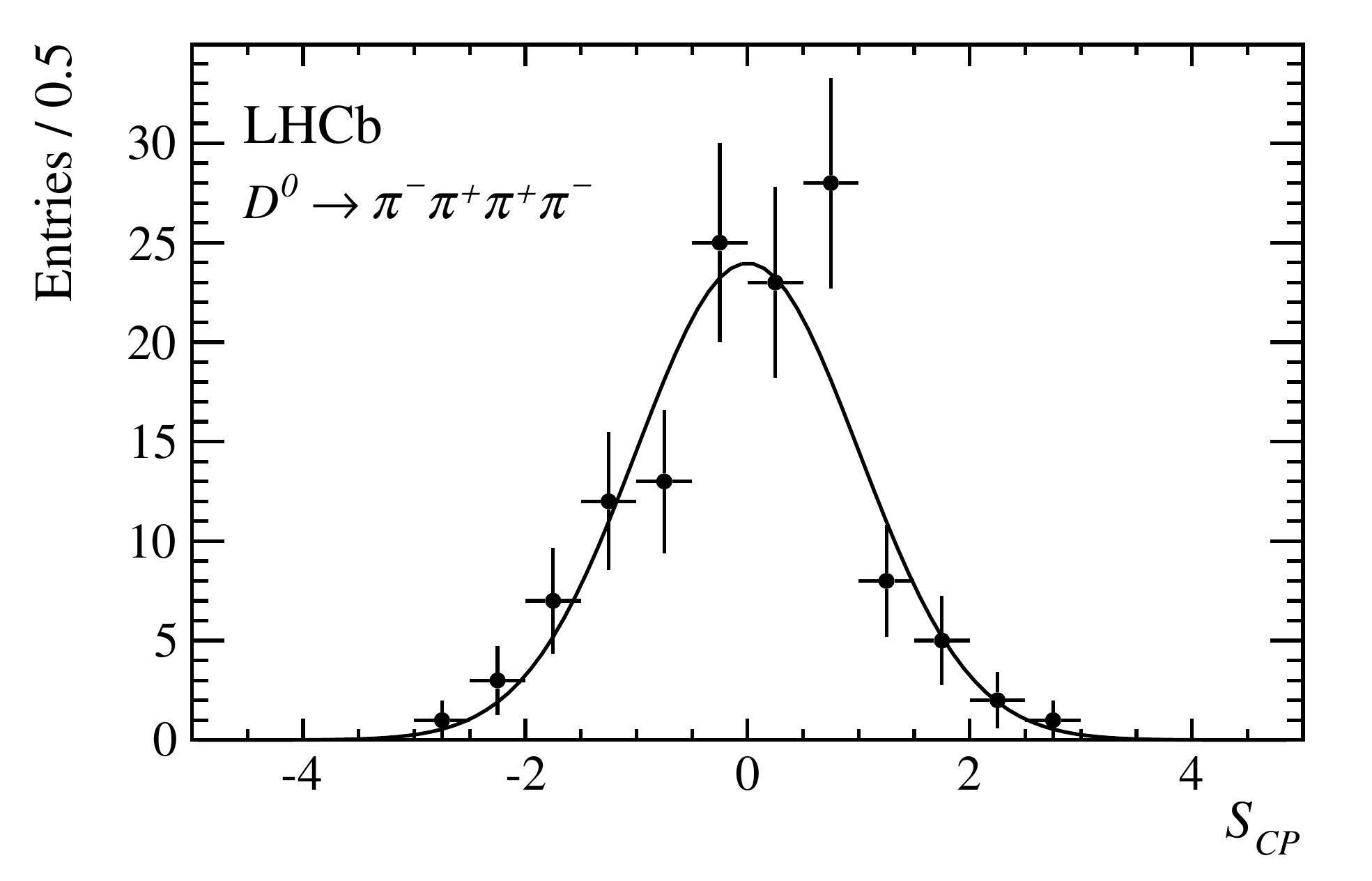}
    \end{minipage}
    \hspace{0.5cm}
    \begin{minipage}[b]{0.48\linewidth}
        \centering
        \includegraphics[width=\linewidth]{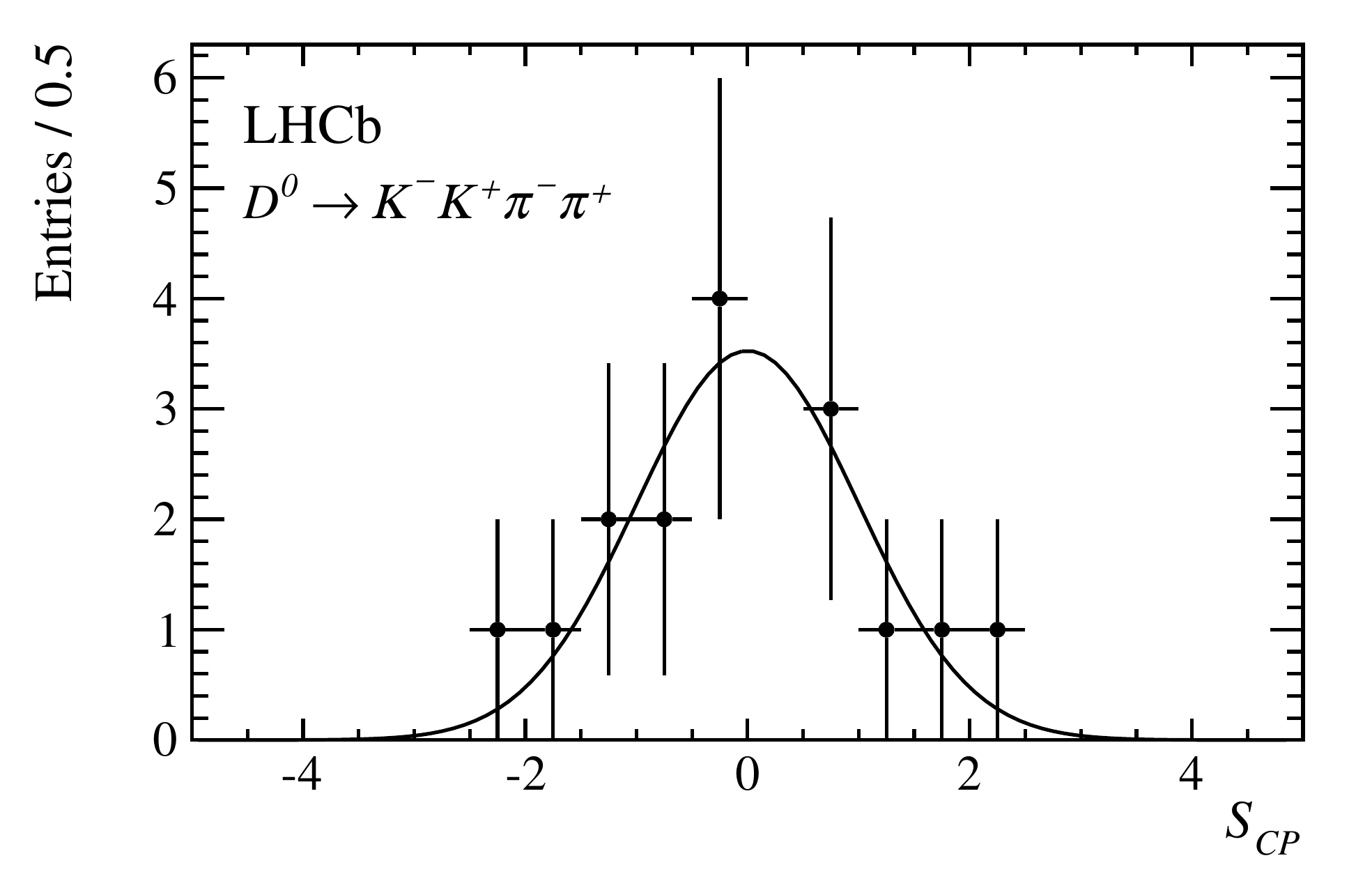}
    \end{minipage}
    \caption{The $S_{CP}$ distribution for (left) $\Dzero \to \pi^{+}\pi^{-}\pi^{+}\pi^{-}$ decays (right) $\DzeroBar \to K^{+}K^{-}\pi^{+}\pi^{-}$ decays, both superimposed with a gaussian distribution of mean 0 and unity width. \label{4bodyscp}}
\end{figure}

The $S^{i}_{CP}$ distribution for both search channels is shown in \figref{4bodyscp}; $D \to K^{-}K^{+}\pi^{-}\pi^{+}$ has a $\chi^2 / \mathrm{NDF} = 42.0/31$ giving a p-value of $9.1\%$, while $D \to \pi^{-}\pi^{+}\pi^{-}\pi^{+}$ has a $\chi^2 / \mathrm{NDF} = 130.0/127$ giving a p-value of $41\%$. Neither shows evidence of local CPV.  

\section{Measuring the $D \to K^{+}\pi^{-}\pi^{+}\pi^{-}$ \catchyName\ using \D mixing }

\label{sec:Mixing}

This section introduces a method that uses \D mixing to constrain the coherence factor and average strong phase difference~\cite{Phen}. Here we discuss application to the final state $K^{+}\pi^{-}\pi^{+}\pi^{-}$ where existing measurements can be considerably improved.  

In the suppressed decay $\Dzero \to K^{+}\pi^{-}\pi^{+}\pi^{-}$ there are two dominant amplitudes; a doubly Cabibbo suppressed (DCS) diagram on one hand, and a time dependant amplitude which proceeds via \D mixing and a Cabibbo favoured (CF) diagram on the other. Having two amplitudes of a comparable magnitude makes this the perfect place to study the interference effects between DCS and CF diagrams. Information on these interference effects are needed to constrain the CP violating phase $\gamma$ in $B^{+} \to DK^{+}$ and similar decay modes. These can be conveniently parameterised by the complex interference parameter \Z~\cite{Phen}, which is related to the coherence factor, \CoherenceFactor, and average strong phase difference, \AveStrongPhaseDiff~\cite{CF, cisi} through $\Z = \CoherenceFactor e^{-i \AveStrongPhaseDiff}$. The magnitude of \Z\ lies in the range $[0,1]$ and gives a measure of how much the interference effects are diluted from integrating over phase-space. The argument of \Z\ gives a weighted average of the strong phase difference between the CF and DCS amplitudes.

In an experimental measurement of the suppressed decay, one usually uses the favoured $\Dzero \to K^{-}\pi^{+}\pi^{-}\pi^{+}$ as a normalisation channel. The theoretical expression for the ratio of decay rates is given by,

\begin{align}
r(t)
 = \frac{R \left ( \Dzero(t) \to f \right )}{R \left (\DzeroBar(t) \to f \right )}  = \rD^{2} + \rD \left( y \ReZ + x \ImZ\right) \gamt +\frac{x^2  + y^2}{4}(\gamt)^{2},  \label{ratio3}
\end{align}
which represents the ratio of rates integrated over all phase space. The dimensionless quantities $x$ and $y$ are the usual \D mixing parameters and $\Gamma$ gives the average width of the \D mass eigenstates. 

A measurement of the constant term in \eqnref{ratio3} allows \rD\ to be constrained, and $\Gamma$ has been well measured previously~\cite{PDG}. Therefore, through the linear term of \eqnref{ratio3} there is sensitivity to $b = y \ReZ + x \ImZ$. 
It is therefore possible to constrain \Z\ given external input on x and y~\cite{HFAG}. These constrains follow a straight line in the $\Imag\Z\ - \Real\Z$ plane which is smeared out by any uncertainty on $x$, $y$ or $b$. 

\begin{figure}[ht]

    \begin{minipage}[b]{0.32\linewidth}
        \centering \small \bf 
        \ \ Toy Simulation
    \end{minipage}
    \begin{minipage}[b]{0.32\linewidth}
        \centering \small \bf
         \ \ Cleo-c
    \end{minipage}
    \begin{minipage}[b]{0.32\linewidth}
        \centering \small \bf
          \ \ Combination
    \end{minipage}

    \begin{minipage}[b]{0.32\linewidth}
        \centering
        \includegraphics[width=\linewidth]{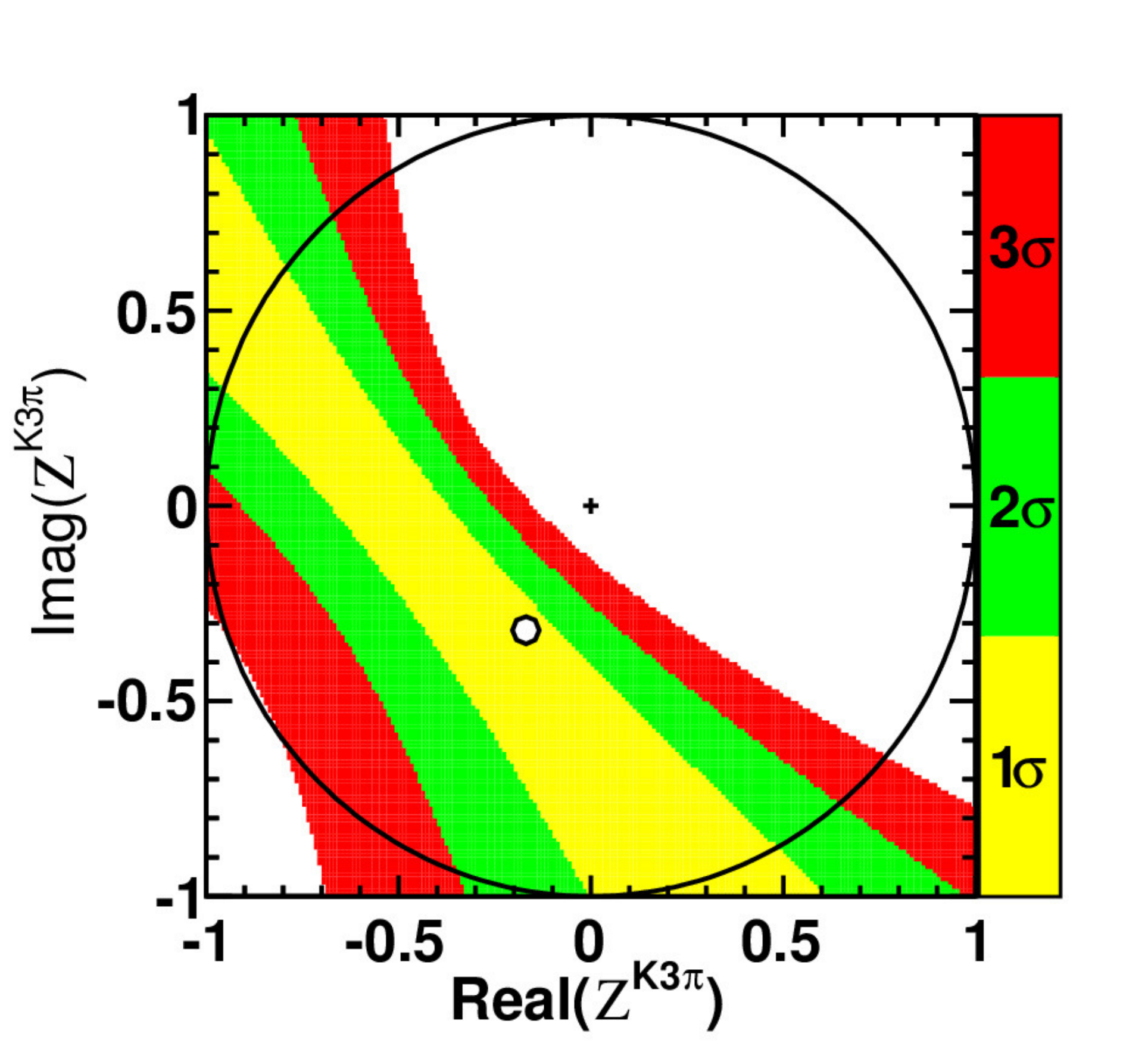}
    \end{minipage}
    \begin{minipage}[b]{0.32\linewidth}
        \centering
        \includegraphics[width=\linewidth]{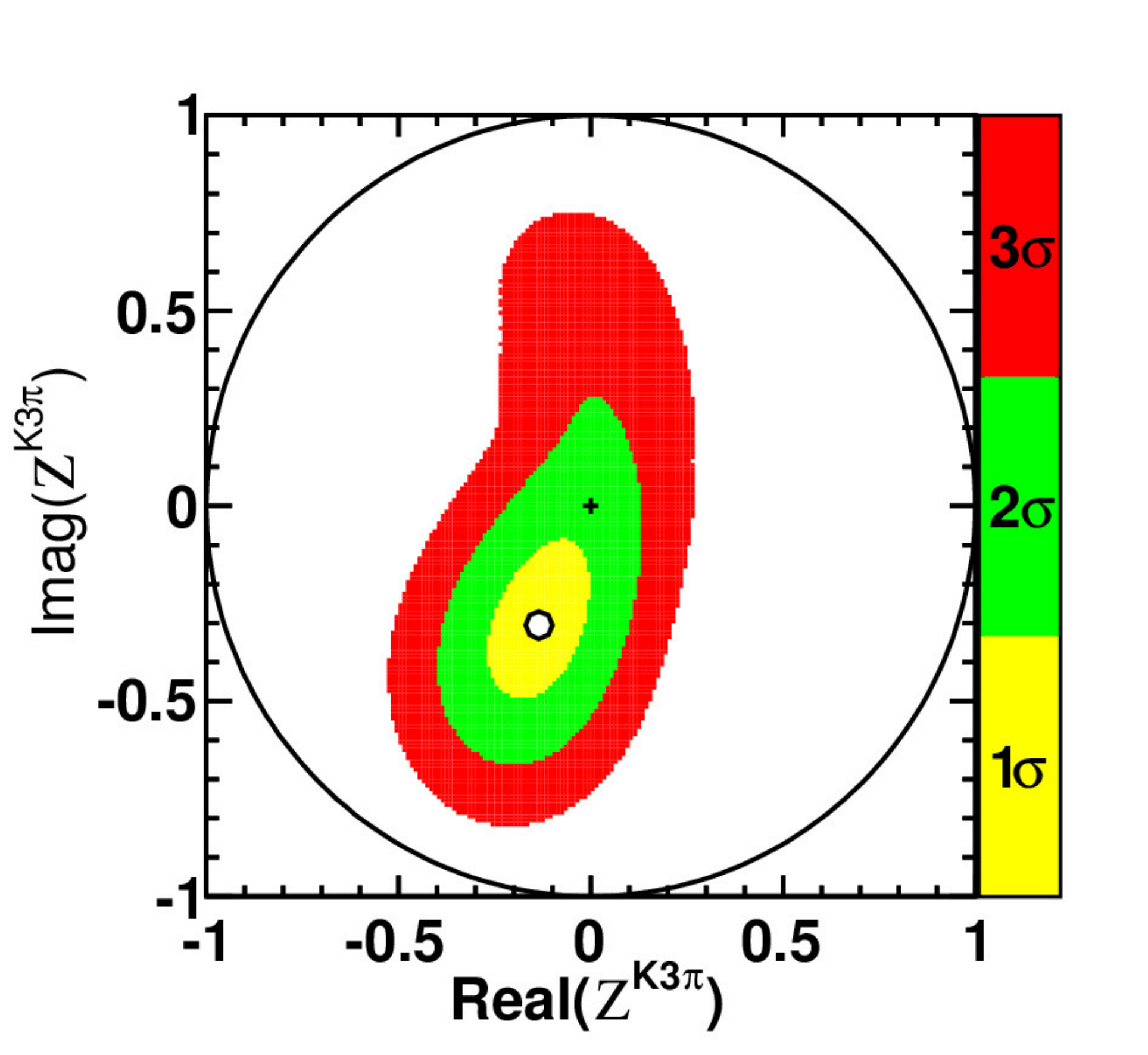}
    \end{minipage}
    \begin{minipage}[b]{0.32\linewidth}
        \centering
        \includegraphics[width=\linewidth]{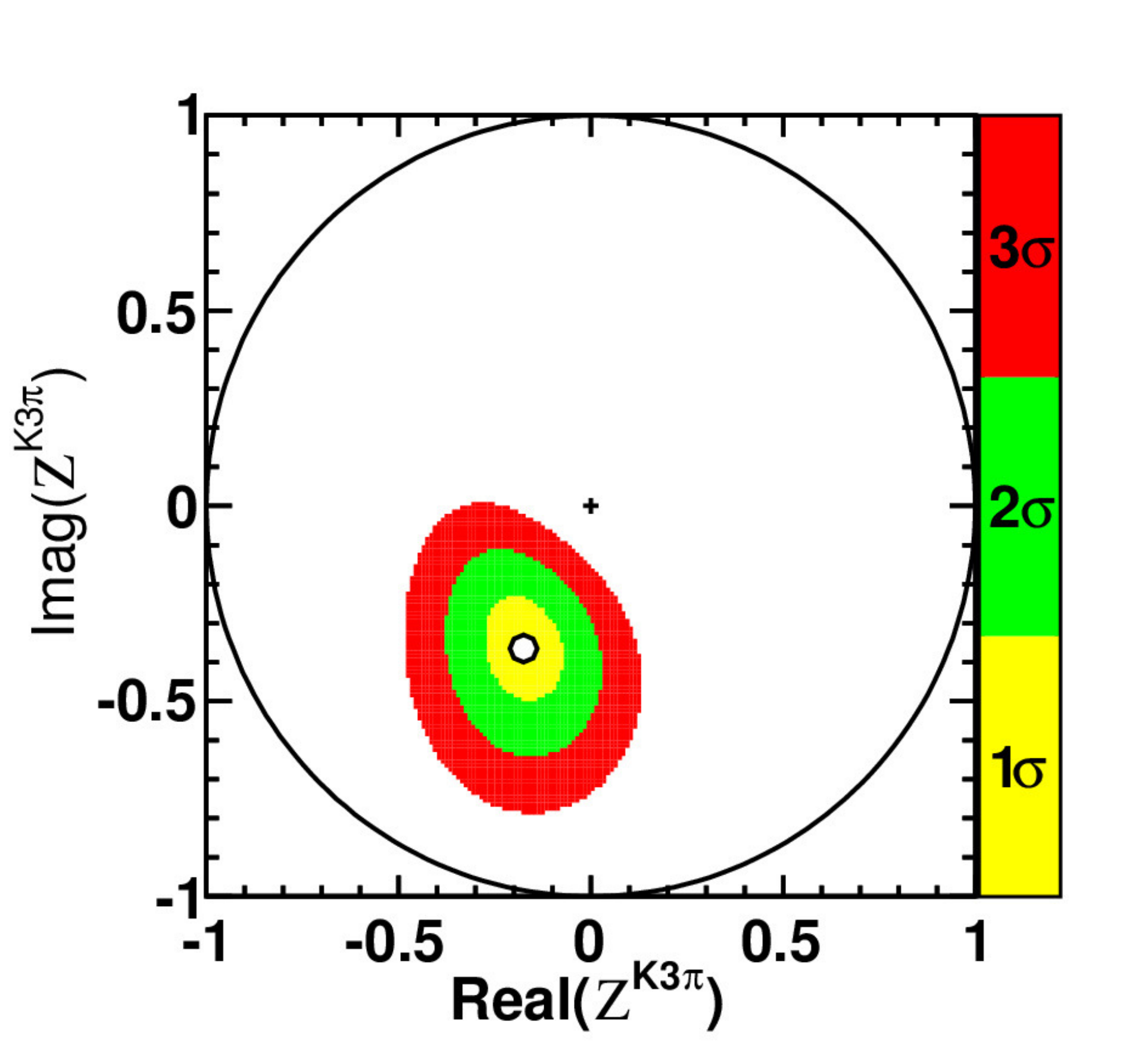}
    \end{minipage}
    \begin{minipage}[b]{0.02\linewidth}
    \centering \small \bf
        \begin{sideways}  \begin{sideways}  \begin{sideways}  Cartesian \ \ \ \ \ \ \ \ \ \end{sideways}  \end{sideways}  \end{sideways}
    \end{minipage}
        
    \begin{minipage}[b]{0.32\linewidth}
        \centering
        \includegraphics[width=\linewidth]{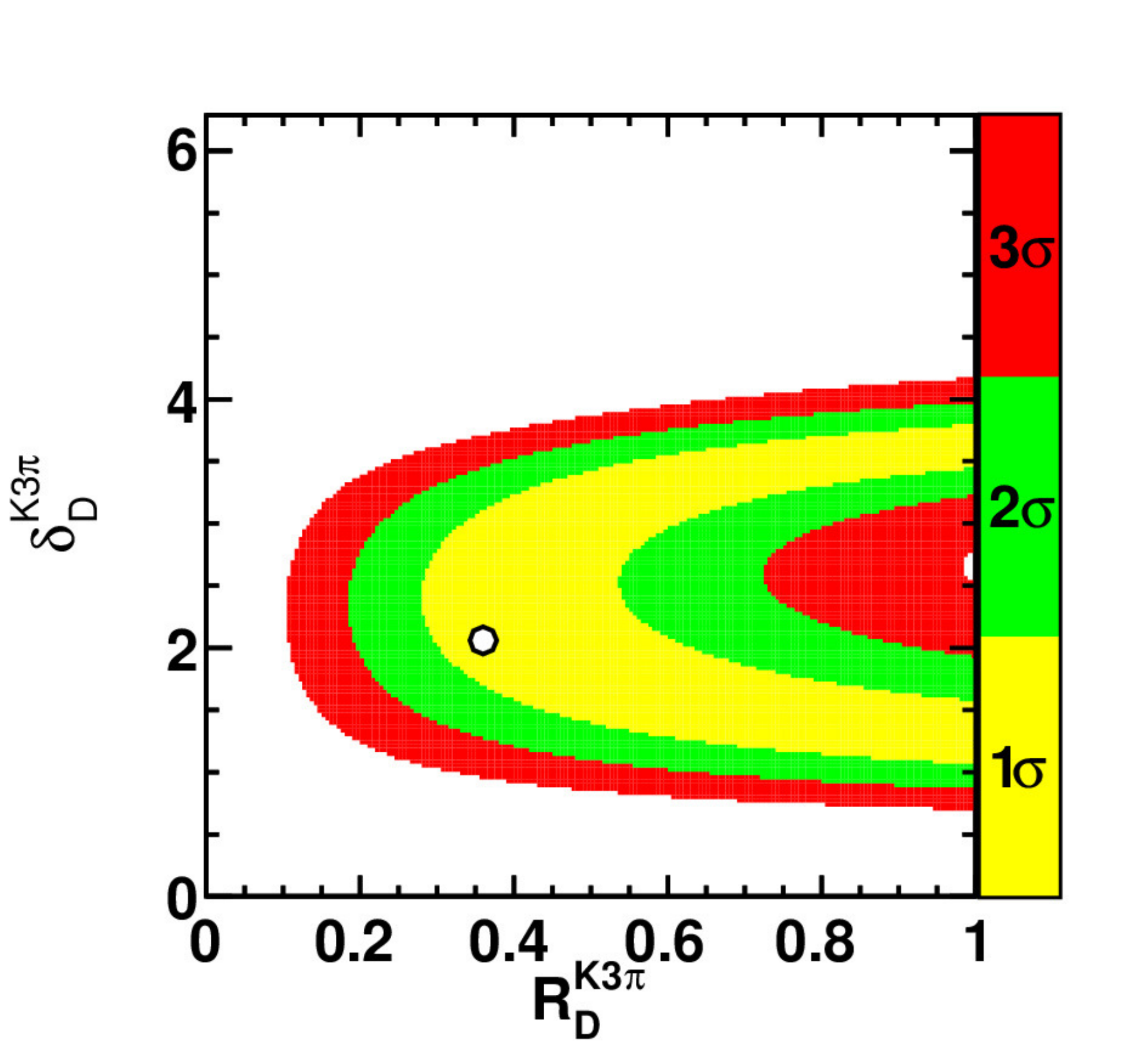}
    \end{minipage}
    \begin{minipage}[b]{0.32\linewidth}
        \centering
        \includegraphics[width=\linewidth]{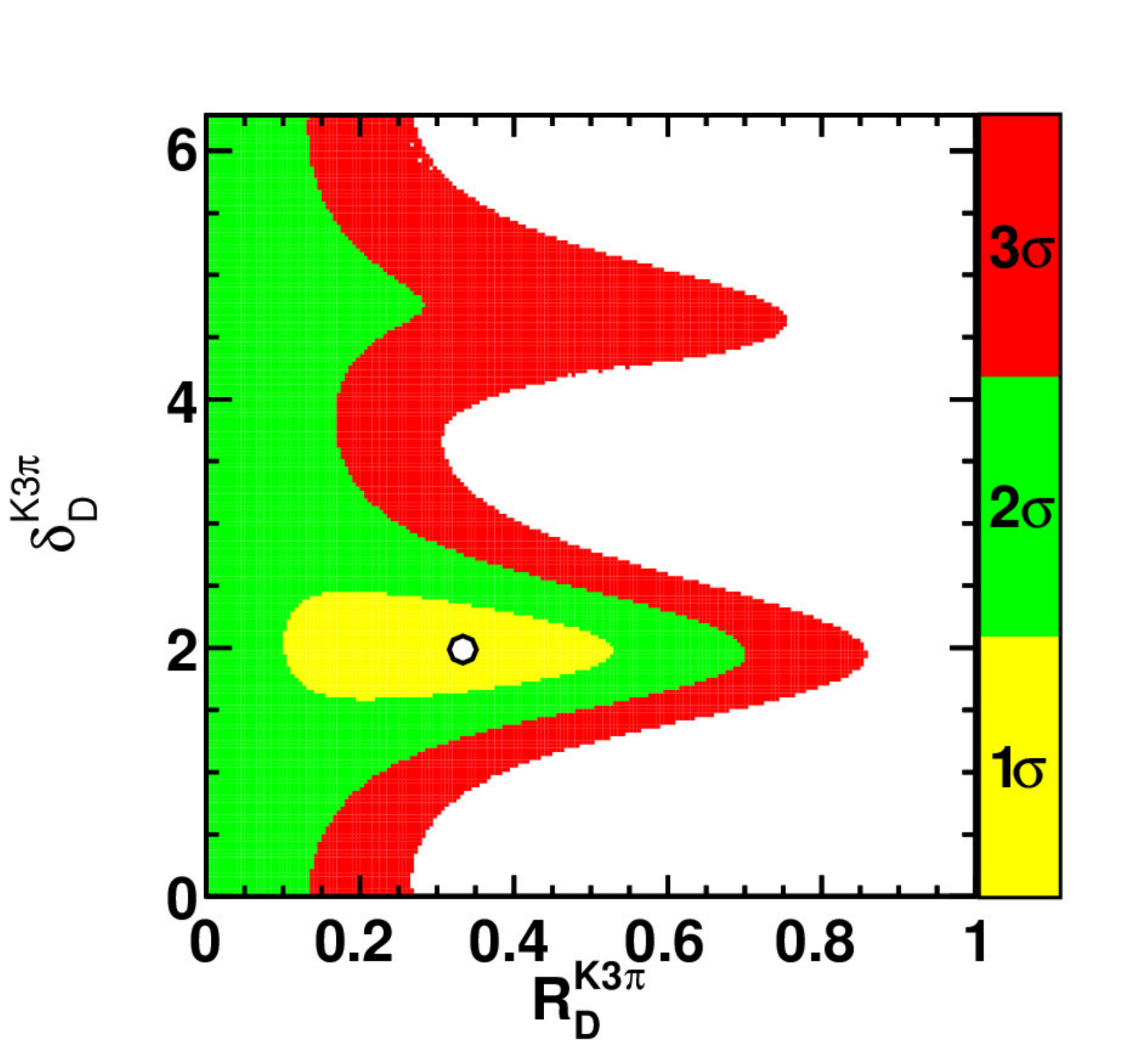}
    \end{minipage}
    \begin{minipage}[b]{0.32\linewidth}
        \centering
        \includegraphics[width=\linewidth]{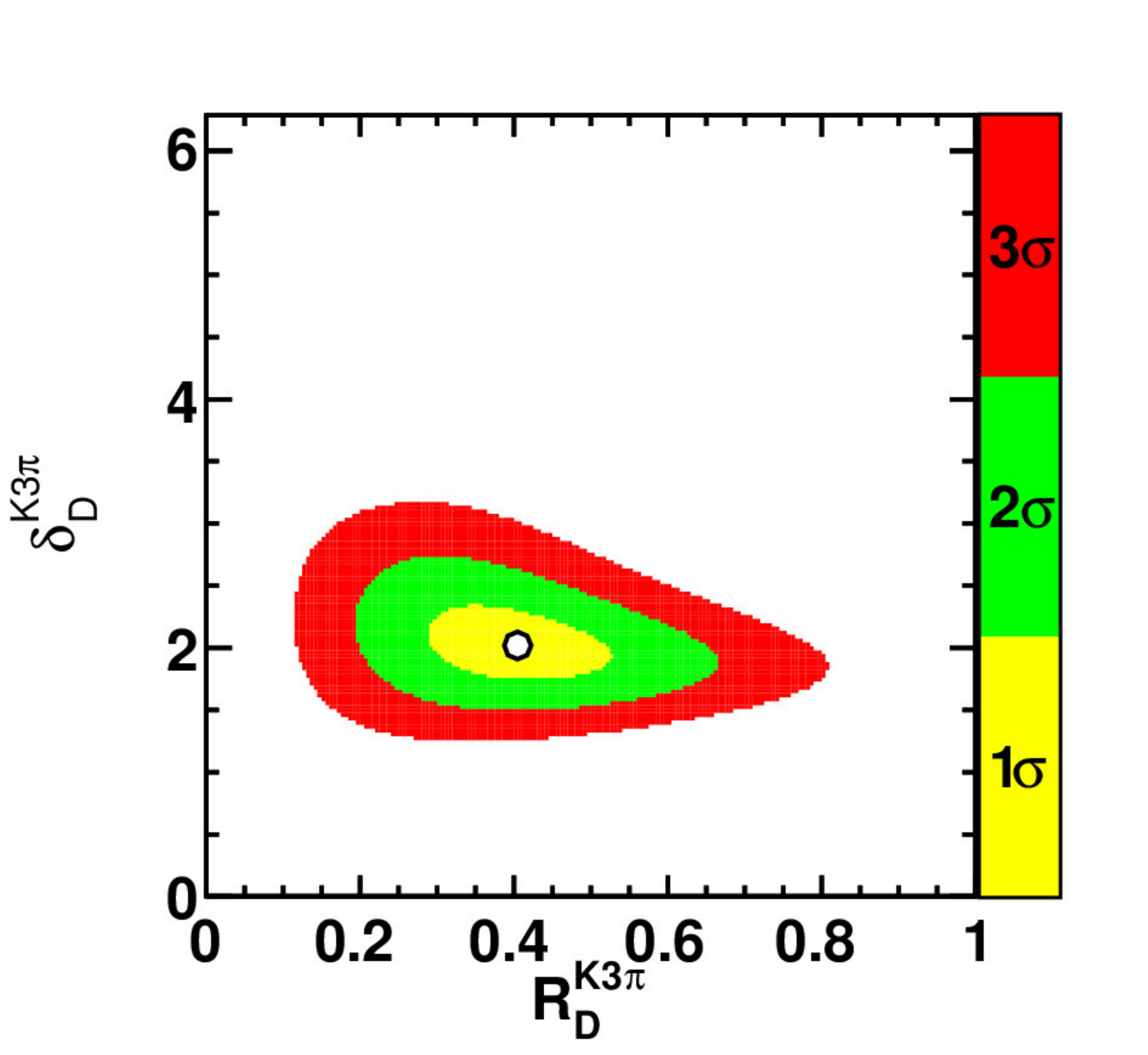}
    \end{minipage}
    \begin{minipage}[b]{0.02\linewidth}
    \centering \small \bf
        \begin{sideways}  \begin{sideways}  \begin{sideways}  Polar \ \ \ \ \ \ \ \ \ \  \ \ \end{sideways}  \end{sideways}  \end{sideways}
    \end{minipage}
   \caption{The figures show constraints on \Z\ from (left) simulated data generated with the Cleo-c central values of \Z\ using expected statistics from LHCb with $3\mathrm{fb}^{-1}$ of data (centre) Cleo-c using threshold data (right) a combination of simulated and threshold data. The top row parameterises the constraints in the preferred \Real \Z\ - \Imag \Z\ coordinates, whereas the bottom row uses the \CoherenceFactor\ - \AveStrongPhaseDiff. \label{fig:chi2scans}} 
\end{figure}

A simulation study based on plausible $\D \to K^{+}\pi^{-}\pi^{+}\pi^{-}$ event yields in LHCb's 2011+2012 dataset leads to the constraints in the \ZKpipipi\ plane shown in \figref{fig:chi2scans}. The results are shown in two separate parameterisations; cartesian coordinates \Real \ZKpipipi\ - \Imag \ZKpipipi\ , and polar coordinates \CoFacKpipipi\ - \AvePDiffKpipipi . 
Also shown in the figure are the constraints set by \cleoc~\cite{CleoCoherence} and a combination of these with the simulated data.
This indicates that considerable improvements on \ZKpipipi\ are possible with currently available datasets.

\section{Conclusions }

Multi-body charm decays have a complex underlying amplitude structure that can lead to localised regions of CP violation across phase space. In $\D$ decays this can give enhanced sensitivity to CPV in charm, a possible signature of new physics. In $B^{+} \to DK^{+}$ decays, one can use the variation of the strong phase to enhance sensitivity to the CP violating phase $\gamma$.

LHCb performed a model independent search for local CPV in $D \to K^{-}K^{+}\pi^{-}\pi^{+}$ and $D \to \pi^{-}\pi^{+}\pi^{-}\pi^{+}$ decays using $1\mathrm{fb}^{-1}$ of data collected in 2011. Assuming there is no CPV, the probability of obtaining the observed results is calculated as $9.1\%$ and $41\%$ for the $D \to K^{-}K^{+}\pi^{-}\pi^{+}$ decay and $D \to \pi^{-}\pi^{+}\pi^{-}\pi^{+}$ decay, respectively. This indicates no evidence for local CPV in either search channel. 

The \catchyName\ plays an important role in measuring the CP violating phase $\gamma$ in $B^{+} \to DK^{+}$ and similar decay modes. Previous constraints on \ZKpipipi, or equivalently \CoFacKpipipi\ and \AvePDiffKpipipi, were set at \cleoc\ using data at the charm threshold. It has been shown that constrains on \catchyName\ can also be found using \D mixing, and a combination of these with existing results could greatly improve the statistical uncertainty on its measurement.


\FloatBarrier








\end{document}

%% file: charmmacros.tex
\newcommand\pubnumber{\pbnr}
\newcommand\pubdate{\today}
\def\Title#1{\begin{center} {\Large #1 } \end{center}}
\def\Author#1{\begin{center}{ \sc #1} \end{center}}

\newcommand{\OnBehalf}[1]{\sbox0{#1}\ifdim\wd0=0pt
        {}
	\else
	{\\on behalf of #1}
	\fi}
\newcommand{\SupportedBy}[1]{\sbox0{#1}\ifdim\wd0=0pt
        {}
	\else
	{\footnote{#1}}
	\fi}
\def\Address#1{\begin{center}{ \it #1} \end{center}}

\newcommand\pubblock{\includegraphics[width=5cm]{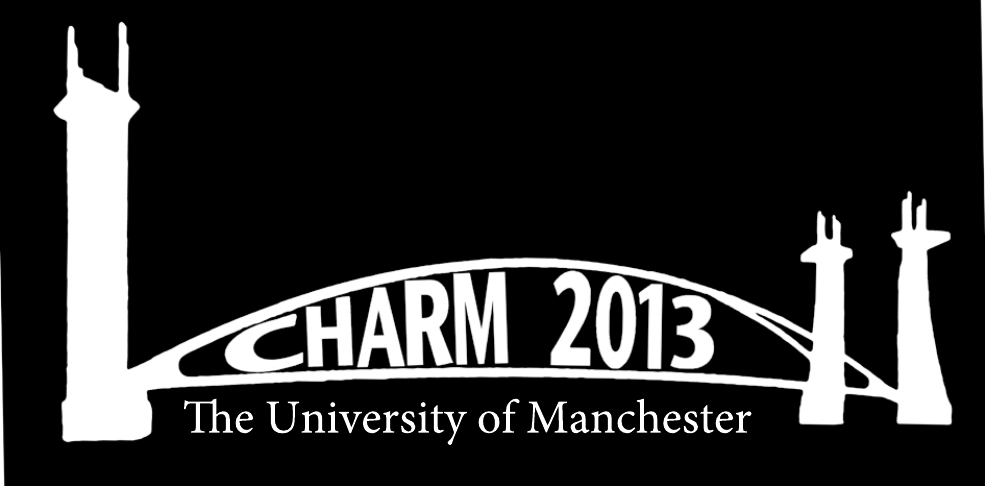}\hfill{\begin{tabular}{l} \pubnumber\\
         \pubdate  \end{tabular}}}
\newenvironment{Abstract}{\begin{quotation}  }{\end{quotation}}
\newenvironment{Presented}{\begin{quotation} \begin{center} 
             PRESENTED AT\end{center}\bigskip 
      \begin{center}\begin{large}}{\end{large}\end{center} \end{quotation}}

\def\venue{The 6$^{th}$ International Workshop on Charm Physics\\
(CHARM 2013)\\
Manchester, UK,  31 August -- 4 September, 2013}

\def\beq{\begin{equation}}
\def\eeq#1{\label{#1}\end{equation}}
\def\eeqn{\end{equation}}

\def\beqa{\begin{eqnarray}}
\def\eeqa#1{\label{#1}\end{eqnarray}}
\def\eeqan{\end{eqnarray}}

\let\bar=\overbar

\def\D{{\cal D}}

\def\Dslash{\not{\hbox{\kern-4pt $D$}}}
\def\dslash{\not{\hbox{\kern-2pt $\del$}}}

\def\ee{e^+e^-}

\def\msb{{\bar{\ssstyle M \kern -1pt S}}}

